# Engaging Stakeholders through Twitter: How Nonprofit Organizations are Getting More Out of 140 Characters or Less

## Kristen Lovejoy; Richard Waters; Gregory D. Saxton

**Abstract**: 140 characters seems like too small a space for any meaningful information to be exchanged, but Twitter users have found creative ways to get the most out of each Tweet by using different communication tools. This paper looks into how 73 nonprofit organizations use Twitter to engage stakeholders not only through their tweets, but also through other various communication methods. Specifically, it looks into the organizations' utilization of tweet frequency, following behavior, hyperlinks, hashtags, public messages, retweets, and multimedia files. After analyzing 4,655 tweets, the study found that the nation's largest nonprofits are not using Twitter to maximize stakeholder involvement. Instead, they continue to use social media as a one-way communication channel, as less than 20% of their total tweets demonstrate conversations and roughly 16% demonstrate indirect connections to specific users.

INTRODUCTION

Launched in October, 2006, Twitter is a short message service, or "micro-blogging" application, that allows users to broadcast real-time messages of 140 characters or less to the entire social media environment.  Since then, Twitter has become the largest micro-blogging site on the Internet. About 19% of all Internet users use Twitter or a similar service for micro-blogging (Fox, Zickuhr, & Smith, 2009), and strategic communicators recognize its ability to reach a large number of stakeholders, as Twitter has become the most-used social media application in official public relations, advertising, and marketing campaigns (Stelzner, 2009).

Social networking sites in general allow not just for the rapid dissemination of information but for the rapid exchange of information. Twitter simplifies this exchange by not only sending real-time messages, but also by limiting the size of the messages, making the information easily digestible. However, some feel that 140 characters is too small a space for any meaningful information to be exchanged and that the popularity of the tool will fade as individuals realize that it is only good for letting the world know what one had for lunch and not for anything more important. Although there may be a great deal of banal chatter on Twitter, organizations are utilizing it for much bigger purposes. The purpose of this paper is to examine how organizations on the "*Nonprofit Times* 100" list use Twitter to engage stakeholders. Specifically, the paper examines how organizations use various communication and interactivity tools specific to Twitter, including following behavior, hyperlinks, hashtags, public messages (PMs), and Retweets (RTs).

LITERATURE REVIEW

*Social Networking and Stakeholder Engagement.*  Social Networking has opened up new possibilities for organizations to engage their stakeholders by allowing them to send information out quickly and to receive real-time feedback.  Most research has focused on the interpersonal implications of social networking; however, the few organizational-level studies point to a great variance in use of social networking to engage stakeholders, with most organizations under-utilizing the technology. Greenberg and MacAulay (2009) found that although some Canadian environmental organizations are fully utilizing the dialogic capacity of social media, most use their sites to simply broadcast messages. Similar results were found for environmental advocacy groups' Facebook pages (Bortree & Seltzer, 2009). Waters, Burnett, Lamm, and Lucas (2009) found similar results for a broader cross-section of the entire nonprofit sector. An examination of the relationship-building features of the 275 randomly-chosen Facebook profiles revealed that these organizations failed to use Facebook for interaction with stakeholders.  All three studies concluded that these organizations had a lost opportunity for furthering dialogue with supporters on Facebook.

The organizational-level research on Twitter is even scarcer. Barnes and Mattson (2010) conducted a study that quantified whether Fortune 500 companies on the 2009 listing had an active Twitter account, being defined as having made an update in the past thirty days. Only 35% of the *Fortune 500* had active accounts. The study further found that 24% of the *Fortune 500* companies actively responded to other users on Twitter and tweeted up-to-date information. One study has examined the different set of priorities between organizations and typical individual users in their use of Twitter (Jansen, Zhang, Sobel, & Chowdury, 2009). The authors found that an organization's Twitter account was "a place for a combination of customer testimony, complaining, feedback, and Q&A" (Jansen, et al., 2009, p. 15). Organizations were less likely to send out "me now"-type tweets, such as what one ate for lunch, and more likely to send out informational messages and to engage their followers with messages that spur dialogue. Overall, Twitter was shown to be a good tool for customer relationship management, though organizations were not fully utilizing it for this purpose (Jansen et al., 2009).

*Communication Tools on Twitter.* While an organization's updates, or tweets, serve as the organization's principal communication tool on Twitter, there are numerous other aspects specific to this micro-blogging application that can aid in stakeholder engagement and organizational research on Twitter. In addition to one-way message announcements, organizations can communicate on Twitter through the use of the "@" symbol. Although Twitter has a way to send direct, private messages, called "direct messages," the norm is to post a tweet with the "@" symbol before the username of the targeted Twitter user. For example, an individual in the dataset asked the March of Dimes, "@marchofdimes Do preemies tend to have higher chances of allergies and sensitivities to food, allergens?" The March of Dimes would see that a user mentioned them in their sidebar and could then reply using the same technique: "@username Preemies have a higher risk for asthma and other lung issues but I haven't found anything about allergies/food sensitivities." Through these public messages (PMs), a dialogue is created between the organization and the user, but it is also viewable by anyone following the March of Dimes' or the individual user's account.

For nonprofits, sending PMs is a way to publicly show responsiveness. PMs are also an important part of creating a dialogue between users and the organization. Users will pose questions and comments to the organization using a PM, and it is important for organizations to respond to and acknowledge these messages. Not responding to a question or comment posed in this way is the equivalent of not responding to an e-mail. It should be noted that some PMs are simply mentions of the organization that do not necessitate a response; however, many still acknowledge users for mentioning the organization. An additional purpose of PMs is to reduce the redundancy of answering the same questions repeatedly in private messages.

An organization that receives many PMs without replying to them would be seen as nonresponsive. However, larger organizations receiving an overwhelming number of PMs could find it overwhelming. For example, one week after the Haitian earthquake, the American Red Cross received more than 900 PMs. Acknowledging each message publicly would not only be time consuming, but would also overload their Twitter updates and might result in users ignoring more relevant messages from the organization. At the opposite end of the spectrum, some organizations are not mentioned by other users at all, which could be a sign the organization is not doing enough to create a dialogue with users. The number of tweets that are PMs has varied in studies from between 12.5% (Java, et.al., 2007) and 22% (Hughes & Palen, 2009). Research has found that the percentage of PMs drops significantly when information-sharing activities increase, such as during major events and crises. PMs accounted for less than 10% of tweets during these events (Hughes & Palen, 2009).

Another common feature used on Twitter is the retweet (RT). This happens when one user simply reposts a tweet from another user and acknowledges the user by adding "RT@[username]" to the beginning of the message. For example, the Make-A-Wish Foundation retweeted the following message from *MacysInc*: "RT @MacysInc: Our Believe stations are overflowing -- 314,000 letters to Santa so far! Don't forget to add yours & help grant WISHES!" Users often solicit others to retweet a post because it contains information that they want shared with a larger audience. Retweets can be used to highlight involvement with another organization, such as in the above example, or to share information that the organization finds pertinent. RTs can also be used when answering PMs, so as to keep the full dialogue together.

The use of hashtags has become common on Twitter to denote that a message is relevant to a particular topic. This makes searching for information easier. For example, if a user wanted to find information about healthcare, a simple search for the term "healthcare" would yield results, but some of these may be off topic; however, a search for "#healthcare" would ensure that all results were relevant to the topic. This communication tool works best when the hashtag has been agreed upon, which usually happens when an organization recommends a specific hashtag to be used by those interested in an event. Hashtags can be vital to getting information out quickly. In wake of the Haitian earthquake, the American Red Cross used the hashtag #Haiti to mark important messages about their relief efforts, and they encouraged individuals to use the hashtag to ask questions about the earthquake's aftermath and spread news about their relief efforts. The use of hashtags can help to sort through information in normal and emergency situations.

Another tool used within tweets are hyperlinks. Many users add links to their website, blog, or other Internet sites to augment the information given in a tweet. Sharing links in a tweet can get followers interested in a story in the same way newspapers use headlines. Organizations encourage followers to read the whole story by following links to non-Twitter websites.

Several third-party websites have been created to help users share information on Twitter. Two popular sites are Twitpic.com, which allows users to link to photos, and Twitvid.com, which allows users to link to videos. Both of these sites point to a growing trend in the use of third-party tools to augment information on Twitter.

Shortened urls, provided by companies such as *bit.ly*, are often used to share these hyperlinks within Twitter's 140-character restriction. URL-shortening services can turn lengthy URLs, such as the Nature Conservacy's http://www.youtube.com/watch?v=Qjg1kMhVvKU (42 characters), into http://bit.ly/3xuuku (20 characters). Character reduction helps conserve space for more pertinent information and attention-seeking headlines.

These *bit.ly* links are interesting to researchers because they can be easily tracked. The *bit.ly* service allows anyone to see how many times a link has been clicked on as well as other information about the link. This allows for a more in-depth analysis of link usage by both the organization and outside researchers.

The aforementioned tools allow organizations to bypass Twitter's 140-character restriction to share a significant amount of information and foster interactivity and engagement with their stakeholders.  But, are nonprofits using these tools to communicate effectively?  The study's research question addresses this concern:

RQ1. Are nonprofit organizations fully utilizing the communication tools available to them on Twitter?

## METHOD

To determine whether Twitter's communication tools were actively being used by nonprofits, a content analysis of organizational tweets was conducted.  Specifically, the sample was taken from the most recent version of the "*Nonprofit Times* 100," which lists the 100 largest non-educational US nonprofit organizations in terms of revenue and is published by *NonProfit Times*. Of these 100 organizations, 73 had Twitter accounts.

Tweets were collected for a month-long period between November 8 and December 7, 2009.  All organizational tweets published during this period were downloaded into an *SQLite* relational database via the Twitter application programming interface using Python code written specifically for this research (available upon request). The final database contained 4,655 tweets, which were doubled checked against the Twitter stream for 10 of the organizations and found to be complete in all cases. The computer-aided content analysis automatically coded for all instances of Twitter's communication tools.  Five percent of the 4,655 sampled tweets were hand coded by two of the researchers to verify computer accuracy; this analysis resulted in an acceptable Cohen's Kappa score for intercoder reliability ($\kappa = .94$) and an accuracy rating of 96.5% for human and computer coding.

## RESULTS

Of the 73 nonprofit organizations featured in the sample, 27% operate in the field of international and foreign affairs, 23% in health, 15% in the arts, culture, and humanities sector, and 8% in youth development, according to their National Taxonomy of Exempt Organizations codes. The remainder of the organizations operated in a variety of areas in the charitable sector. On average, the nonprofit organizations that were examined followed an average of 2,842 users ($sd = 6,946.8$), though this ranged considerably from a high of 46,723 to a low of 3.  The organizations sent more than two tweets per day during the month ($m = 66.23$, $sd = 65.74$), though this varied significantly as the number of tweets ranged from 0 to 289.  Proportionally, these tweets were more likely to contain hyperlinks (68%) or hashtags (29.9%) than they were to contain messages that were public replies (16.2%) or retweets (16.2%). However, there is great variation in the use of each of these functions by the individual organizations as well as their overall behavior on Twitter.

*Following on Twitter.*  At its core, Twitter is a micro-blogging site; however, it also functions as a social networking site in that users can connect and share information.  Using Twitter terminology, if one user follows another, he/she is considered a "follower," and if both users follow each other they are considered "friends." To examine the friending practices of nonprofit organizations, the researchers set up a Twitter account that was used to follow all 73 sampled organizations. The researchers' Twitter accounts ultimately were followed by only 17 of the 73 organizations in the sample. Since the account was created solely for data collection and not for engagement, it can be deduced that these organizations systematically follow anyone who follows them. Following users that follow an organization gives the impression that the organization wants to know what those who are interested in the organization are talking about, even if they never actually read this information. Creating mutual ties with followers is one way organizations can at least give the appearance of creating a community on Twitter. Conversely, an organization that does not follow anyone gives the impression that they do not want to engage in a dialogue.

The organizations in this sample followed as few as 3 and as many as 46,723 users, showing the large variance in friending behavior. Young Life, which followed the fewest users, only follows users that it finds may help advance its mission.  While the strategy of only following those that are deemed interesting or helpful may be a typical strategy for individual users, it is not conducive to organizational relationship building. Some organizations, such as the New York Public Library, only follow users that are affiliated with them, such as New York Public Library Kids.

*Tweets.*  The sample of organizations sent out a total of 4,655 tweets over the November 8 – December 7 time period. The nonprofits used Twitter consistently throughout the month. The average number of tweets per organization for the first two weeks (m = 33.38, sd = 35.50) were roughly half of that of the entire month (m = 66.23, sd = 65.74).

The frequency with which an organization sends out tweets is used to consider how active an organization is. This is in line with prior research, where "active" individual users have been categorized as those who post once or more per week and "inactive" users as those who posted less than once a week (Hughes & Palen, 2009). Organizations are held to a different standard than individual users in terms of activeness. A study by Sysmos showed that users who self-identified as social media marketers are far more active, 6.3% post two updates a day and 4.3% post at least nine updates a day (Cheng, Evans, & Singh, 2009). Followers expect organizations, like other social media marketers, to be more active than the average individual user. To determine whether the nonprofits were active Twitter users, the researchers examined each account to see if at least 3 tweets were sent over the first two weeks of the study. Of the 73 organizations, 80.8% (n = 59) organizations were classified as active.  If an organization is tweeting fewer than 3 times per week, its tweets may get buried in its followers' feeds. However, sending out too many tweets may clutter its followers' feeds and result in users un-following the organization.

*Hyperlinks.*  The majority of the nonprofits' tweets (n = 3,170) included hyperlinks to external information.  At 68% of the total, the usage of hyperlinks by the nonprofits is considerably greater than that of the average individual user on Twitter, which has been estimated to be between 13% and 25% (Java et al., 2007; Hughes & Palen, 2009). Organizations are more official information sources than individuals, which explains the greater proportion of links to information subsidies. Only one organization in the sample did not use any hyperlinks in its tweets.

In the sample, 21 organizations used Twitpic.com to send a total of 61 links to photos, and only one used Twitvid.com to link to a video. The lack of popularity of these sites in this sample may be due to the fact that all of these organizations have their own websites, which they can use to post pictures and videos, whereas most individual users on Twitter do not.

*Public Messages.*   Organizations were often found to send informative links to all of their Twitter followers as well as specific individuals using public messages (PM). Of the 4,655 tweets made by the nonprofits during the month examined, 16.2% (n = 756) of the total were PMs, characterized as any message that started with the "@" symbol.  Further analysis of the data revealed that 16 organizations in the sample (21.9%) received no PMs from other Twitter users.

*Retweets.*  Nonprofits in the sample used the retweet (RT) function less often than Twitter users in general.  Hughes and Palen (2009) estimated that individuals use the function 27.8% of the time.  The nonprofits in this sample had 755 tweets that simply shared other users' tweets. This represents 16.2% of the total number of the nonprofits' tweets during the month.  There was considerable variation among the organizations.  Two organizations used the function more than 50 times, while 15 never used it.  The majority fell in the middle, with 24 of the organizations sending out 10 or more RT messages.

*Hashtags.*  Nearly 30% of the nonprofits' tweets (n = 1,394) included one or more hashtags.  Reflecting widespread variation of this communication tool, 11 nonprofits never used a hashtag within their tweets, and 10 organizations used hashtags more than 40 times.  Three organizations (CARE, American Cancer Association, and World Vision) were statistical outliers as they used hashtags 120, 160, and 259 times, respectively.  The use of hashtags by these organizations is likely a sign that the nonprofits have a better understanding of how searches occur on Twitter and focus more on search engine optimization than the others in the sample.

DISCUSSION

Despite the "140-character" restriction, the results of the current study reveal that Twitter is a more complex communication tool than might be expected at first glance. Savvy organizations are able to bypass the character restriction to present detailed information through the use of hyperlinks, to construct replies to public messages that demonstrate responsiveness to constituent concerns, to facilitate rapid diffusion of information by retweeting messages, to build information communities and aid in Twitter searches by using hashtags, and to share multimedia files by using the TwitPic and TwitVid services.  The results of this study reflect similar public relations studies examining how organizations use social media.  Information dissemination in the form of sharing hyperlinks and retweeted messages were the two dominant communication tools used by the sampled nonprofit organizations.

For the past several years, consultants have been pushing for public relations to adopt social media to grow virtual communities with stakeholders (e.g., Li & Bernoff, 2008; Solis & Breakenridge, 2009); however, there have been only minimal results that indicate social media is becoming a mainstay in public relations programming.  Certainly, the usage is on an uptick, as would be expected with the introduction of any new communication channel, but there has been little evidence of interactivity and relationship-building.  Instead, the current study continues to reveal that organizations, in this case the nation's largest nonprofits, are continuing to use Twitter as they would a traditional information subsidy.

Just as scholars have found that Facebook failed to capitalize on the engagement elements of the site (e.g., Bortree & Seltzer, 2009; Waters, Burnett, Lamm, & Lucas, 2009) and that blogs are predominantly one-way message channels (e.g., Seltzer & Mitrook, 2007), Twitter is proving to be yet another social media outlet being hyped for relationship-building efforts that public relations practitioners do not fully perceive as being present.  Rather than using public messages to reply to other Twitter users or connecting to others by retweeting messages that may be helpful to others, nonprofits are primarily using the site to relay information using one-way communication.

One has to wonder why public relations practitioners are not using the interactive elements in the proportions that they are advocated by consultants.  Kent (2008) cautioned organizations to venture carefully into social media, especially blogs, as little evidence exists that it truly can build communities around organizations, and responsiveness to blog postings are limited to a small handful of individuals.  Additionally, research has shown that individuals are very apathetic to organizations' use of social media, as they themselves primarily use the services to connect with friends, family, and co-workers (Vorvoreanu, 2008).  But perhaps the lack of time and resources being put into organizations' social media accounts stems from the lack of research indicating that social media use produces support for short-term or long-term financial benefits for the organization (Hearn, Foth, & Gray, 2009).

There may be a more simplistic reason for the lack of interaction on organizations' social media accounts.  Despite the suggestions by consultants, practitioners may neither understand nor believe that social media is the cure-all for organizational communication efforts.  Social media consultants reiterate the power of social media by focusing on customer service issues (e.g., Li & Bernoff, 2008; Solis & Breakenridge, 2009), but practitioners have had a history of battling organizational perceptions equating public relations and customer service.

The Excellence Theory found that public relations thrived when it was recognized as a management counseling function, not a lower-level function putting out individual fires (J. Grunig, L. Grunig, & Dozier, 1992), and these findings are reiterated throughout public relations management and strategy textbooks as well as the leading research journals.  This study's findings help echo these concerns as communication tools that promote interactivity are also used in customer service situations.  Perhaps practitioners are facing cognitive dissonance from being told by consultants that they should embrace a lower-level customer service function rather than engaging in traditional boundary spanning and environmental scanning for the management consulting level.  No doubt, social media can be used to accomplish both tasks; however, consultants and "how-to" handbooks have yet to recognize this key difference.

Until the field decides which direction to pursue, research will most likely continue to produce results similar to this study's findings.  Although some of the nonprofit organizations in the sample are using Twitter to create a real dialogue, most are still using it as just another way to send out information such as that found in traditional newsletters, media kits, and annual reports.  Although it may seem counterintuitive that real interactions can happen in 140 characters or less, Twitter can be used as a tool for stakeholder engagement—if practitioners take the initiative to use it proactively to meet the traditionally taught boundary spanning and environmental scanning roles of the discipline's management function, or to use it reactively for customer service as suggested by consultants.

CONCLUSION

While Twitter is the leading social media outlet for organized campaign efforts, strategic communicators still remain puzzled over how to best use Twitter to connect with their external stakeholders on a daily basis (Stelzner, 2009).  As found in this study, organizations are only limited in how they use Twitter by the imaginations of their public relations practitioners.  While many may perceive 140-characters restrictive in the amount and type of information that can be shared, Twitter offers a variety of communication tools that allow organizations to bypass the reliance on short messages.  The current findings indicate that organizations vary significantly in the way that they use the different tools.  However, a few limitations should be discussed before concluding about how nonprofits use Twitter.

*Limitations.*  The nonprofits chosen for the study were selected using an established list of the largest nonprofits in the United States.  While this choice was made for its provision of a sound design framework, it limits the understanding of organizational use of Twitter.  Smaller, community-based nonprofits were excluded from the research in favor of large, national nonprofits.  It can be argued that smaller, grassroots nonprofits may be more interactive and use conversational tweets with their followers rather than using one-way information dissemination practices.  Another limitation of the study is that it cannot measure the number of private direct messages made by the nonprofits.  While the proportion of public messages was relatively low and indicated that the nonprofits may not be engaging with other Twitter users frequently, this conclusion may not be true.  Private messages between organizations and users are not available for analysis.  Perhaps users' questions or concerns were addressed privately rather than through the public message function; this private messaging would demonstrate conversation, and it might be a preferred method of relationship-building because of its one-on-one communicative nature.  One final limitation concerns the research method.  While content analysis reveals how the communication tools are being used, it does not measure the underlying motivations that practitioners have for using them.  Understanding strategic communicators' attitudes toward the various communication tools and Twitter may provide more insights into organizations' use and views toward the impact of social media application.

*Future Research.*  In addition to future research addressing the limitations, additional research needs to be conducted to analyze other dimensions of Twitter usage.  For example, comparisons between the nonprofit, for-profit, and government sectors may reveal variations in how the various sectors communicate with its Internet audiences.  Additionally, the number of communication tools has changed since the data for this study were collected.  Twitter now allows users to create lists of users, which could aid organizational communication to specific groups, and allows users to share their geographic location with their tweets, which organizations could use to attract audiences to specific events.  In September, 2010, Twitter announced a major upgrade to the site that allows easier sharing of multimedia files, which could further help organizations share information.  However, perhaps one of the most beneficial studies would examine how Twitter is used in connection with other Internet sites to build organizational communities.  Are the hyperlinks shared on Twitter sending users to blogs, online petitions, surveys, or Facebook content?  Exploring where these hyperlinks connect would reveal insights into how organizations view social media's role in organizational endeavors.  Is it primarily used as an outlet to collect research on stakeholders, or is it to disseminate information in a one-way manner?  Social media consultants stress that Twitter and other social media are the channel practitioners should be focused on for relationship cultivation with stakeholders, yet this study continues the string of research showing that organizations have not fully embraced this function.